\documentclass[10pt,letterpaper]{article}
\usepackage{opex3}
\usepackage{graphicx,psfrag,epsf,epsfig}% Include figure files
\usepackage{dcolumn}% Align table columns on decimal point
\usepackage{bm,bbm}% bold math
\usepackage{pstricks,pst-plot}     
\usepackage{latexsym}
\usepackage{mathrsfs,amsmath,amssymb,textcomp}
\usepackage{caption}
\usepackage{hyperref}
\newcommand{\pfq}[1]{\frac{\partial^{2}}{\partial^{2} #1}}

\begin{document}

%%%%%%%%%%%%%%%%%% title page information %%%%%%%%%%%%%%%%%%
\title{Simulation of light propagation in thin semiconductor films with non-local electron-photon
interaction}

\author{U. Aeberhard}

\address{IEK5-Photovoltaik, Forschungszentrum J\"ulich, \\ D-52425 J\"ulich, Germany}

\email{u.aeberhard@fz-juelich.de} %% email address is required



\begin{abstract}
The propagation of light in layered semiconductor media is described theoretically and
simulated numerically within the framework of the non-equilibrium Green's function formalism as used
for state-of-the-art nanodevice simulations, treating the non-local interaction of leaky photonic modes
with the electronic states of thin semiconductor films on a non-equilibrium quantum statistical
mechanics level of theory. For a diagonal photon self-energy corresponding to local coupling, the
simulation results for a 500 nm GaAs slab under normal incidence are in excellent agreement with the
predictions from the conventional transfer matrix method. The deviations of the local
approximation from the result provided by the fully non-local photon self-energy for a 100 nm GaAs
film are found to be small.
\end{abstract}
 
% REPLACE WITH CORRECT OCIS CODES FOR YOUR ARTICLE

%%%%%%%%%%%%%%%%%%%%%%% References %%%%%%%%%%%%%%%%%%%%%%%%%
\bibliographystyle{osajnl}
%\bibliography{/home/aeberurs/Biblio/bib_files/negf,/home/aeberurs/Biblio/bib_files/aeberurs,/home/aeberurs/Biblio/bib_files/scqmoptics,/home/aeberurs/Biblio/bib_files/generation,/home/aeberurs/Biblio/bib_files/optical_modelling,/home/aeberurs/Biblio/bib_files/lecture_notes,/home/aeberurs/Biblio/bib_files/recombination}

%%%%%%%%%%%%%%%%%%%%%%%%%%  body  %%%%%%%%%%%%%%%%%%%%%%%%%%
\section{Introduction}

Many novel architectures for opto-electronic devices such as solar cells or light emitting 	diodes
utilize nanostructures as functional components for enhanced device performance. These nanoscale 
components interact with the electromagnetic radiation on a length scale in the sub-wavelength
regime. 	In these devices, complex combinations of dielectric and electronic nanostructures lead to
an intricate pattern of interaction of leaky photonic modes with confined electronic states.
While adequate descriptions 	of electronic transport in structures with strong spatial inhomogeneity at the
nanoscale, 	such as the non-equilibrium Green's function formalism (NEGF) \cite{datta:95}, are well
established 	and routinely used, the consideration of the interaction of charge carriers with photon
modes is in most cases still based on the local coupling to classical solutions of Maxwell's
equations. 	In order to fully account for electronic coherence and optical confinement in the 
description of the electron-photon interaction for photocurrent generation, a non-local theory 	of
photon modes in optically open devices is required, which is compatible with the description of
charge carrier transport.

In this paper, such a description is provided based on the photon Green's functions (GF) for general 
non-equilibrium conditions and one dimensional spatial composition variation of the optically active
medium. After a brief review of the general NEGF theory of optical modes, the steady-state equations
are formulated 	for a semiconductor slab system and subsequently solved numerically for normal
incidence of polychromatic light and both local and non-local coupling to the electronic system.

\section{NEGF formalism for optical modes}
The optical NEGF problem of slab systems has been discussed in the literature for several
different applications, such as emission enhancement in microcavity lasers
\cite{jahnke:95}, absorption-reflection set-up for a general non-equilibrium system
\cite{richter:08,henneberger:09,henneberger:09pra} or the photovoltaic response of a
spatially homogeneous (bulk) absorber and in absence of recombination losses \cite{mozyrsky:07}. 
Here, the focus is on the absorption of external radiation that is coupled into the same leaky modes
that 	are responsible fo the emission of light leading to radiative dark current. This treatment is 
required for a consistent microscopic theory of nanostructure based solar cell devices 
\cite{ae:jcel_11,ae:jstqe_13}, in which the electron-photon coupling responsible for photocarrier
generation is described in terms of a self-energy including the (non-local) GFs of
both charge carriers and photons.

 \subsection{Photon Green's function}
The photon GF can be defined via Maxwell's equation for the effective vector
potential of the electromagnetic field $\mathbf{A}_{eff}(\underbar{1})\equiv \langle 
\hat{\mathbf{A}}(\underbar{1})\rangle_{C}$ ($\underbar{1}\equiv \{\mathbf{r}_{1},\underbar{t}_{1}\in
C\}$, $C$: Keldysh contour \cite{keldysh:65}), which can be written as a free field expansion in
terms of bosonic operators,
\begin{align}
\hat{\mathbf{A}}({\mathbf r},\underbar{t})=&\sum_{\lambda,{\mathbf
q}}\left[\mathbf{A}_{0}(\lambda,\mathbf{q}) \hat{b}_{\lambda,{\mathbf
q}}(\underbar{t})e^{i{\mathbf q}{\mathbf r}}+\mathbf{A}_{0}^{*}(\lambda,\mathbf{q}) 
\hat{b}_{\lambda,{\mathbf q}}^{\dagger}(\underbar{t})e^{-i{\mathbf q}{\mathbf
r}}\right]\label{eq:photfieldop},\\
\mathbf{A}_{0}(\lambda,\mathbf{q})=&\sqrt{\frac{\hbar}{2\epsilon_{0}V\omega_{\lambda\mathbf{q}}}}
{\mathbf \epsilon}_{\lambda{\mathbf q}},
\end{align}
and induced current $\mathbf{j}_{ind}(\underbar{1})\equiv \langle \hat{\mathbf{j}}(\underbar{1})\rangle_{C}$, which then reads
\begin{align}
\left(\Delta-\frac{n(\mathbf{r})}{c_{0}^{2}}\pfq{\underbar{t}}\right)\mathbf{A}_{eff}(\mathbf{r},\underbar{t})=
-\mu_{0}[\mathbf{j}_{ind}(\mathbf{r},\underbar{t})+\mathbf{j}_{ext}(\mathbf{r},\underbar{t})]
\end{align}
for an external current density $\mathbf{j}_{ext}$ (c-number function). The photon GF is then defined via
the functional derivative\footnote{SI units are used throughout the paper.}
\begin{align} 
\mathcal{D}_{\mu\nu}(\underbar{1},\underbar{2})=&-\frac{1}{\mu_{0}}\frac{\delta
A_{eff,\mu}(\underbar{1})}{\delta j_{ext,\nu}(\underbar{2})}\\=&
-\frac{1}{\mu_{0}}\frac{i}{\hbar}\left[\langle \hat{A}_{\mu}(\underbar{1})
\hat{A}_{\nu}(\underbar{2})\rangle_{C}-A_{eff,\mu}(\underbar{1})A_{eff,\nu}(\underbar{2})\right],
\end{align}
where $\mu_{0}=1/(\varepsilon_{0}c_{0}^2)$ is the magnetic vacuum permeability. Similarly, the transverse 
polarization function corresponding to the photon self-energy is defined via
\begin{align}
\Pi_{\mu\nu}(\underbar{1},\underbar{2})=&-\mu_{0}\frac{\delta
j_{ind,\mu}(\underbar{1})}{\delta A_{eff,\nu}(\underbar{2})}.
\end{align}
The contour ordered photon GF follows as the solution of the corresponding
Dyson equation\footnote{We assume Einstein's convention of
summation over repeated indices.}
\begin{align}
\int
d3\left[\mathcal{D}_{0,\mu\beta}^{-1}(\underbar{1},\underbar{3})-\Pi_{\mu\beta}(\underbar{1},\underbar{3})\right]
\mathcal{D}_{\beta\nu}(\underbar{3},\underbar{2})
=\delta_{\parallel,\mu\nu}(\underbar{1},\underbar{2}),
\end{align}
where $\mathcal{D}_{0,\mu\nu}$ is the free propagator defined by 
\begin{align}
\mathcal{D}_{0,\mu\nu}^{-1}(\underbar{1},\underbar{2})=\left[\Delta_{1}-\frac{1}{c^2}\frac{\partial^2}{\partial
\underbar{t}_{1}^2}\right]\delta_{\mu\nu}\delta(\underbar{1},\underbar{2}),
\end{align}
and $
\delta_{\parallel,\mu\nu}(\underbar{1},\underbar{2})=\delta(\underbar{t}_{1}-\underbar{t}_{2})\delta_{\parallel,\mu\nu}(\mathbf{r}_1-\mathbf{r}_2)$
is the transverse delta function,
\begin{align}
\delta_{\parallel,\mu\nu}(\mathbf{r}_1-\mathbf{r}_2)=&\delta_{\mu\nu}\delta(\mathbf{r}_1-\mathbf{r}_2)
+\nabla_{\mu}\nabla_{\nu}\frac{1}{4\pi|\mathbf{r}_1-\mathbf{r}_2|}\\
=&\frac{1}{(2\pi)^3}\int
d^{3}k\left(\delta_{\mu\nu}-\frac{k^{\mu}k^{\nu}}{k^2}\right)e^{i\mathbf{k}(\mathbf{r}_1-\mathbf{r}_2)}.\label{eq:transdelt}
\end{align}
The corresponding real time components of the retarded photon GF are determined
via the Dyson equation
\begin{align}
\int
d3\left[\mathcal{D}_{0,\mu\beta}^{-1}(\underbar{1},\underbar{3})-\Pi_{\mu\beta}^{R}(\underbar{1},\underbar{3})\right]
\mathcal{D}_{\beta\nu}^{R}(\underbar{3},\underbar{2})=&\delta_{\parallel,\mu\nu}(\underbar{1},\underbar{2}).\label{eq:Dyson}
\end{align}
The kinetic or Keldysh equations for the photon correlation functions read
\begin{align}
\int
d3\left[\mathcal{D}_{\mu\beta}^{R,-1}(\underbar{1},\underbar{3})
\mathcal{D}_{\beta\nu}^{\lessgtr}(\underbar{3},\underbar{2})
-\Pi_{\mu\beta}^{\lessgtr}(\underbar{1},\underbar{3})\mathcal{D}_{\beta\nu}^{A}(\underbar{3},\underbar{2})\right]
=&0.
\end{align}
Finally, the photon spectral function is defined via
\begin{align}
\hat{\mathcal{D}}_{\mu\nu}(\underbar{1},\underbar{2})=i[\mathcal{D}_{\mu\nu}^{R}(\underbar{1},\underbar{2})
-\mathcal{D}_{\mu\nu}^{A}(\underbar{1},\underbar{2})].\label{eq:photspecfun}
\end{align}
In analogy to the electronic case, we define the local photonic density of states at steady-state
($t_{1}-t_{2}=\tau\rightarrow E$) via
\begin{align}
\mathcal{N}(\mathbf{r};E)=\frac{\mathcal{C}}{2\pi}\mathrm{Tr}\{\hat{\boldsymbol{\mathcal{D}}}(\mathbf{r},\mathbf{r};E)\}
\label{eq:photlocdos}
\end{align}
where the normalization factor $\mathcal{C}=\mu_{0}/(A_{0}^{2}V)=2E/(\hbar
c_{0})^2$ ($E=\hbar\omega$) takes account of the relation of the the photon spectral
function to the spectral function of non-interacting bosons \cite{derlet:96}.

\subsection{Quasi-1D model for layer structures}
In layer structures with homogeneous transverse dimensions, the steady-state equations for the
photon GFs can be simplified by using the Fourier transform of the latter with respect to transverse coordinates, i.e.,
\begin{align}
\mathcal{D}_{\mu\nu}(\mathbf{r},\mathbf{r}',E)=\frac{A}{(2\pi)^2}\int
d^{2}q_{\parallel}\mathcal{D}_{\mu\nu}(\mathbf{q}_{\parallel},z,z',E)
e^{i\mathbf{q}_{\parallel}\cdot(\mathbf{r}_{\parallel}-\mathbf{r}'_{\parallel})},
\end{align}
with  $A$ the cross-section area. For each energy and transverse momentum vector, a
separate set of equations for the GFs needs to be solved. In the case of the retarded
GF, moving the free propagator to the right in Eq.
\eqref{eq:Dyson} leads to the integral form of the Dyson equation for the dyadic form,
\begin{align}
\mathcal{D}_{\mu\nu}^{R}(\mathbf{q}_{\parallel},z,z',E)=&\mathcal{D}_{0\mu\nu}^{R}(\mathbf{q}_{\parallel},z,z',E)\nonumber\\&+\int
dz_{1}\int
dz_{2}\mathcal{D}_{0\mu\alpha}^{R}(\mathbf{q}_{\parallel},z,z_{1},E)\Pi_{\alpha\beta}^{R}(\mathbf{q}_{\parallel},z_{1},z_{2},E)
\mathcal{D}_{\beta\nu}^{R}(\mathbf{q}_{\parallel},z_{2},z',E).\label{eq:dyson}
\end{align}
Similarly, the kinetic equation for the correlation functions becomes
\begin{align}
\mathcal{D}_{\mu\nu}^{\lessgtr}(\mathbf{q}_{\parallel},z,z',E)=&\int dz_{1}\int
dz_{2}\mathcal{D}_{\mu\alpha}^{R}(\mathbf{q}_{\parallel},z,z_{1},E)\Big[
\Pi_{0\alpha\beta}^{\lessgtr}(\mathbf{q}_{\parallel},z_{1},z_{2},E)\nonumber\\&+
\Pi_{\alpha\beta}^{\lessgtr}(\mathbf{q}_{\parallel},z_{1},z_{2},E) \Big]
\mathcal{D}_{\beta\nu}^{A}(\mathbf{q}_{\parallel},z_{2},z',E),\label{eq:keldysh}
\end{align}
where the self-energy components related to the solution of the \emph{homogeneous} problem, i.e.,
incident fluctuations that are independent from the state of the absorber, are given
by \cite{mozyrsky:07,richter:08}
\begin{align}
\Pi_{0\mu\nu}^{\lessgtr}(\mathbf{q}_{\parallel},z,z',E)=\int dz_{1}\int
dz_{2}[\mathcal{D}_{0}^{R}]^{-1}_{\mu\alpha}(\mathbf{q}_{\parallel},z,z_{1},E)
\mathcal{D}_{0\alpha\beta}^{\lessgtr}(\mathbf{q}_{\parallel},z_{1},z_{2},E) 
[\mathcal{D}_{0}^{A}]^{-1}_{\beta\nu}(\mathbf{q}_{\parallel},z_{2},z',E)
\end{align}
in terms of the GFs of the unperturbed system.

In the case of a one-dimensional dielectric perturbation potential
(i.e., a 1D photonic crystal), the unperturbed GFs $\mathcal{D}_{0}$ can be defined on
the basis of solutions for a homogeneous free space, which in the general case are given
by (see App. \ref{sec:appa})
\begin{align}
\mathcal{D}_{0\mu\nu}(\mathbf{q},E)=\frac{\hbar c_{0}^2}{2V}\sum_{\lambda}\frac{{\mathbf 
\epsilon}_{\lambda{\mathbf q}}^{\mu}{\mathbf\epsilon}_{\lambda{\mathbf
q}}^{\nu}}{\omega_{\lambda\mathbf{q}}}D_{0\lambda}(\mathbf{q},E),
\end{align}
where $D_{0}$ is the scalar  GF of non-interacting bosons. In the case of a
homogeneous medium (e.g., vacuum), the retarded component of the free GF does not
depend on polarization, and the polarization sum can be performed explicitly,
\begin{equation}
\sum_{\lambda}\epsilon^{\mu}_{\lambda{\mathbf
q}}\epsilon^{\nu}_{\lambda{\mathbf
q}}=\delta^{\mu\nu}-\frac{q^{\mu}q^{\nu}}{q^2}\equiv\delta^{\parallel}_{\mu\nu}(\mathbf{q}),
\label{eq:polprod}
\end{equation}
where $\delta^{\parallel}(\mathbf{q})$ is the transverse delta function in reciprocal space as given
in Eq. \eqref{eq:transdelt}. The same applies to the correlation functions, if the occupation of
the modes is polarization independent, such as in the case of unpolarized incident light or
completely isotropic internal emission. 

The polarization-averaged GFs in slab representation are then obtained via inverse Fourier
transform. Inserting the explicit forms for the GF components of non-interacting bosons in
homogeneous systems (bulk) yields the expressions ($V=A\cdot L$)
\begin{align}
\mathcal{D}^{R}_{0\mu\nu}(\mathbf{q}_{\parallel},z,z',E)=&\frac{\hbar^2c_{0}^2}{A}\int
\frac{dq_{z}}{2\pi}\delta^{\parallel}_{\mu\nu}(\mathbf{q})e^{i
q_{z}(z-z')}\left[(E+i\eta)^2-(\hbar\omega_{q})^2\right]^{-1},\\
\mathcal{D}^{\lessgtr}_{0\mu\nu}(\mathbf{q}_{\parallel},z,z',E)=&-\frac{i\hbar^2c_{0}^2}{2A}\int
dq_{z}\delta^{\parallel}_{\mu\nu}(\mathbf{q})e^{\pm
iq_{z}(z-z')}(\hbar\omega_{q})^{-1}\nonumber\\&\times \Big[N^{0}_{{\mathbf q}}\delta(E\mp\hbar\omega_{q})+(N^{0}_{-{\mathbf
q}}+1)\delta(E\pm\hbar\omega_{q})\Big],
\end{align}
where $q=|\mathbf{q}|=\sqrt{q_{\parallel}^2+q_{z}^2}$ and $\hbar\omega_{q}=\frac{c_{0}}{n_{0}}\hbar q$. Due to the
$\mathbf{q}$-dependence of $\delta^{\parallel}$, the free GF can be separated in isotropic and
anisotropic contributions,
\begin{align}
\mathcal{D}_{0\mu\nu}=\mathcal{D}^{(1)}_{0}\delta_{\mu\nu}+\mathcal{D}^{(2)}_{0\mu\nu}.
\end{align}
In the case of the retarded component,  further evaluation requires consideration of the polarization components. 
The scalar isotropic term (corresponding to the \emph{Huygens} propagator) is obtained as
\begin{align}
\mathcal{D}^{R(1)}_{0}(\mathbf{q}_{\parallel},z,z',E)&=-\frac{in_{0}^2}{2A}\frac{\exp\left[iq_{z0}(q_{\parallel},E)|z-z'|\right]}
{q_{z0}(q_{\parallel},E)},
\end{align}
where $q_{z0}(q_{\parallel},E)=\sqrt{q_{0}^2-q_{\parallel}^2}$ with $q_{0}=\frac{n_{0}}{\hbar
c_{0}}|E|$.
Including the full anisotropy corresponds to the consideration \emph{dyadic} propagator
\cite{keller:11} 
\begin{align}
\mathcal{D}^{R}_{0}(\mathbf{q}_{\parallel},z,z',E)=&-\frac{in_{0}^2}{2A}\frac{\exp\left[iq_{z0}(q_{\parallel},E)|z-z'|\right]}
{q_{z0}(q_{\parallel},E)q_{0}^{2}}\Big\{q_{0}^{2}\mathbbm{1}-\mathbf{q}_{\parallel}\otimes\mathbf{q}_{\parallel}
-[q_{z0}(q_{\parallel},E)]^{2}\mathbf{\hat{z}}\otimes\mathbf{\hat{z}}\nonumber\\&-\left(
\mathbf{q}_{\parallel}\otimes\mathbf{\hat{z}}+\mathbf{\hat{z}}\otimes\mathbf{q}_{\parallel}\right)[q_{z0}(q_{\parallel},E)]^{2}\mathrm{sgn}(z-z')\Big\}.
\end{align}
In the above expression, the first term in the curly bracket is the isotropic part
$\mathcal{D}^{R(1)}_{0}$, while the remaining terms constitute the anisotropic part
$\mathcal{D}^{R(2)}_{0}$. In the evaluation of the correlation function for free-field modes, the
directional dependence of the occupation number needs to be considered. Transforming the $\delta$-function in the energy
domain to the corresponding expression in terms of $q_{z}$ results in (for $E>0$ and omitting the
momentum and energy dependence of $q_{z0}$ for clarity)
\begin{align}
\mathcal{D}^{<}_{0\mu\nu}(\mathbf{q}_{\parallel},z,z',E)&=-\frac{in_{0}^{2}}{2A}\sum_{\sigma=+,-}
\delta^{\parallel}_{\mu\nu}(\mathbf{q}_{\parallel},q_{z0\sigma})\exp[iq_{z0\sigma}(z-z')]q_{z0}^{-1}
N^{0}_{\mathbf{q}_{\parallel},q_{z0\sigma}},\label{eq:corrfn_free_in}\\
\mathcal{D}^{>}_{0\mu\nu}(\mathbf{q}_{\parallel},z,z',E)&=-\frac{in_{0}^{2}}{2A}\sum_{\sigma=+,-}
\delta^{\parallel}_{\mu\nu}(\mathbf{q}_{\parallel},q_{z0\sigma})\exp[-iq_{z0\sigma}(z-z')]q_{z0}^{-1}
(N^{0}_{-\mathbf{q}_{\parallel},-q_{z0\sigma}}+1),\label{eq:corrfn_free_out}
\end{align}
where $q_{z0\sigma}=\sigma q_{z0}$.
\subsection{Photon self-energy}

For a simple 1D dielectric potential, the retarded photon self-energy reduces to
the diagonal term
$\Pi_{\mu\nu}(\mathbf{q}_{\parallel}=0,z,z',E)=V(z,E)\delta_{\mu\nu}\delta(z-z')$, where the
dielectric potential is given by
\begin{align}
V(z,E)=-\big[n(z,E)^2-n_{0}^2\big]q_{0}^{2},
\end{align}
with $n(z,E)$ the spatially varying refractive index at given photon energy, while
$n_{0}$ and $c_{0}$ are refractive index and speed of light in vacuum. The effect of photon
absorption can be 	considered for both local and non-local interaction. In the local case, the
refractive index 	in the expression for the perturbation potential entering the diagonal photon
self-energy is 	replaced by a complex value including the extinction coefficient $\kappa$: 
$n(z,E)=n_{r}(z,E)+i\kappa(z,E)$. 	For the consideration of the non-local interaction as mediated,
e.g., by the electron-photon 	self-energy within the NEGF formalism of photogeneration
\cite{ae:prb_11}, the full photon 	self-energy needs to be evaluated,
\begin{align}
\Pi_{\mu\nu}^{\alpha}(\mathbf{q}_{\parallel},z,z',E)=-i\hbar\mu_{0}\Big(\frac{e}{m_{0}}\Big)^{2}p_{cv}^{\mu*}(z)
\mathcal{P}_{cv}^{\alpha}(\mathbf{q}_{\parallel},z,z',E) p_{cv}^{\nu}(z'),\qquad
\alpha=\lessgtr,R,\label{eq:photse}
\end{align}
where $\mathbf{p}_{cv}$ is the (local) momentum matrix and the interband polarization function is
given by
\begin{align}
 \mathcal{P}_{cv}^{\gtrless}(\mathbf{q}_{\parallel},z,z',E)=&A^{-1}\sum_{\mathbf{k}_{\parallel}}\int 
 \frac{dE'}{2\pi\hbar}G_{cc}^{\gtrless}(\mathbf{k}_{\parallel},z,z',E')G_{vv}^{\lessgtr}(\mathbf{k}_{\parallel}
 -\mathbf{q}_{\parallel},z',z,E'-E),\label{eq:polfun}\\
  \mathcal{P}_{cv}^{R}(\mathbf{q}_{\parallel},z,z',E)=&A^{-1}\sum_{\mathbf{k}_{\parallel}}\int 
 \frac{dE'}{2\pi\hbar}\Big[G_{cc}^{R}(\mathbf{k}_{\parallel},z,z',E')G_{vv}^{<}(\mathbf{k}_{\parallel}
 -\mathbf{q}_{\parallel},z',z,E'-E)\nonumber\\
 &+G_{cc}^{<}(\mathbf{k}_{\parallel},z,z',E')G_{vv}^{A}(\mathbf{k}_{\parallel}
 -\mathbf{q}_{\parallel},z',z,E'-E)\Big].
\end{align} 
For local coupling to classical fields, the local extinction coefficient for given angle of
incidence and polarization is related to the non-local photon self-energy via the local absorption
coefficient \cite{ae:unpublished},
\begin{align}
\kappa_{\mu}(\mathbf{q}_{\parallel},z,E)=\alpha_{\mu}(\mathbf{q}_{\parallel},z,E)\cdot\frac{\hbar
c_{0}}{2E}=\frac{(\hbar c_{0})^{2}}{4 n_{r}E^{2}}\int dz'
\Re[i\Pi_{\mu\mu}^{>}(\mathbf{q}_{\parallel},z',z,E)].\label{eq:extcoef}
\end{align}

\subsection{Physical quantities}
Together, the different components of the photon GF provide all the spectral and integral
observables that can be described in terms of single-photon operator averages. Most relevant for
opto-electronic device applications such as solar cells or light emitting diodes are the local
density of photon states, the local photon density and the local value of the photon flux or
Poynting vector. According to \eqref{eq:photlocdos}, the general expression for the local density of
photon states (LDOS) is found from the retarded GF as follows:
\begin{align}
 \mathcal{N}(z,E)=&-\frac{\mathcal{C}}{\pi}\sum_{\mu}\sum_{\mathbf{q}_{\parallel}}
 \mathrm{Im}\mathcal{D}_{\mu\mu}^{R}(\mathbf{q}_{\parallel},z,z,E)\equiv
 \frac{\mathcal{C}}{2\pi}\sum_{\mu}\sum_{\mathbf{q}_{\parallel}}
 \hat{\mathcal{D}}_{\mu\mu}(\mathbf{q}_{\parallel},z,z,E).\label{eq:ldos}
\end{align}
The local photon density is obtained from the
correlation function, 
\begin{align}
n_{\gamma}(\mathbf{q}_{\parallel},z,E)=\frac{\mathcal{C}}{2\pi}\sum_{\mu}i\mathcal{D}_{\mu\mu}^{<}(\mathbf{q}_{\parallel},z,z,E).
\label{eq:photdens}
\end{align}
Finally, the  $z$-component of the modal Poynting vector is given by \cite{richter:08}
\begin{align}
s_{z}(\mathbf{q}_{\parallel},z,E)=&-\frac{E}{2\pi\hbar}\lim_{z'\rightarrow
z}\partial_{z'}\sum_{\mu}\mathrm{Re}\Big[\mathcal{D}_{\mu\mu}^{>}(\mathbf{q}_{\parallel},z,z',E)+
\mathcal{D}_{\mu\mu}^{<}(\mathbf{q}_{\parallel},z,z',E)\Big].\label{eq:poyntvec}
\end{align}

\section{Analytical and numerical verification for normal incidence}
\begin{figure}[tb]
\begin{minipage}{0.475\textwidth}
\includegraphics[width=\textwidth]{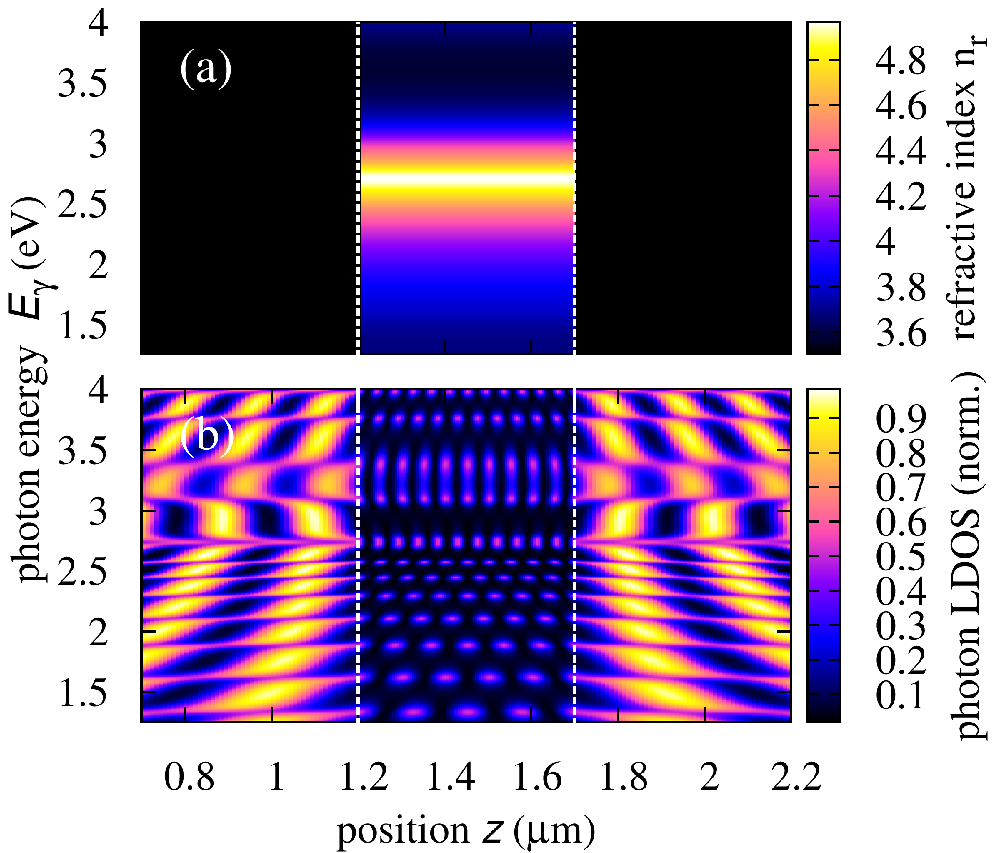}
\captionof{figure}{(a) Refractive index profile and (b) corresponding local density of photon
 states at vanishing transverse photon momentum $\mathbf{q}_{\parallel}=0$ for a 500 nm GaAs
 slab.}\label{fig:ldos}
\end{minipage}
\hspace{0.05\textwidth}
\begin{minipage}{0.475\textwidth}
\includegraphics[width=\textwidth]{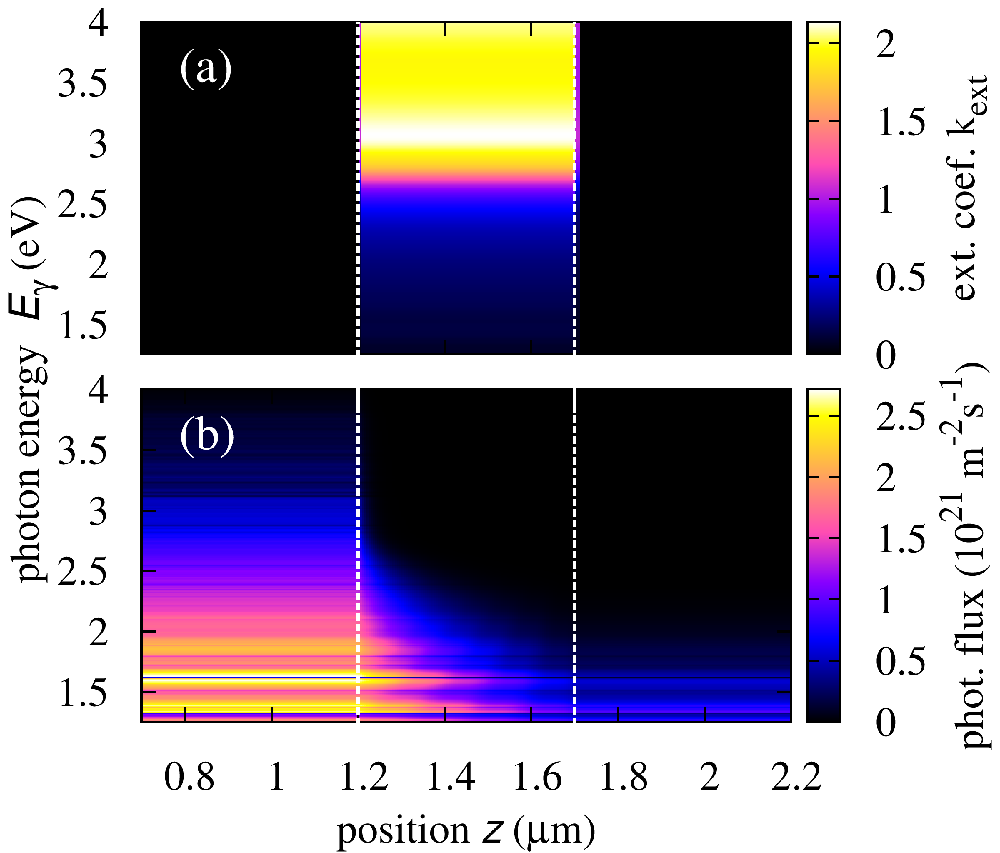}
\captionof{figure}{(a) Extinction coefficient $\kappa$ and (b) spectral photon flux for the AM1.5
solar spectrum under normal incidence ($\mathbf{q}_{\parallel}=0$).\newline}  \label{fig:flux}
\end{minipage}
\end{figure}
As a first consistency check, the local photon density of states is computed, which in the case of
free field modes should be independent of space and amount to $\mathcal{N}_{0}(z,E)=(n_{0}^3
E^2)/(\pi^2\hbar^3 c_{0}^3)$.  Evaluating the different expressions for the
diagonal polarization components for the retarded GF results in an isotropic contribution of (see App.
\ref{sec:appb})
\begin{align}
\mathcal{N}_{0}^{(1)}(z,E)=3\cdot \frac{n_{0}^3 E^2}{2\pi^2\hbar^3 c_{0}^3}
\end{align}
and an anisotropic contribution of
\begin{align}
\mathcal{N}_{0}^{(2)}(z,E)=-\frac{n_{0}^3 E^2}{2\pi^2\hbar^3 c_{0}^3},
\end{align}
which together provides the above photon DOS of free field modes. 

To verify the computation of the GF for an inhomogeneous situation and local coupling, the local
photon DOS, spectral density of photons and local photon flux for a 500 nm thick GaAs slab in
air are compared to the corresponding quantities as computed via a standard transfer matrix method
(TMM), using the same optical data (SOPRA database) and the standard AM1.5 solar spectrum at
normal incidence as external photon source. In the case of 	normal incidence
($\mathbf{q}_{\parallel}=0)$, the anisotropic part vanishes in $\mathcal{D}_{0}$. The corresponding
integral 	equation for the scalar component of the retarded photon can be solved 	numerically either
via quadrature \cite{kahen:93} or a recursive approach \cite{kahen:93,rahachou:05}. Due to the
nonlocality 	of the electron-photon coupling, we adopt the former 	approach here, since it provides 
the full GF including all off-diagonal elements. It amounts to the 	linear problem (at fixed energy $E$)
\begin{align}
\pmb{\mathcal{M}}\cdot\pmb{\mathcal{D}}^{R(1)}=\pmb{\mathcal{D}}^{R(1)}_{0},
\end{align}
with 
\begin{align}
\mathcal{M}_{ij}=\delta_{ij}-\mathcal{D}_{0}^{R(1)}(z_{i},z_{j})V(z_{j}).
\end{align}  

The LDOS [Fig.
\ref{fig:ldos}(b)] and photon density are evaluated for a real refractive index profile [Fig.
\ref{fig:ldos}(a)], i.e., neglecting absorption, while the photon flux [Fig. \ref{fig:flux}(b)] is
computed under inclusion of the extinction coefficient [Fig. \ref{fig:flux}(a)].
\begin{figure}[t]
\begin{center}
\includegraphics[width=\textwidth]{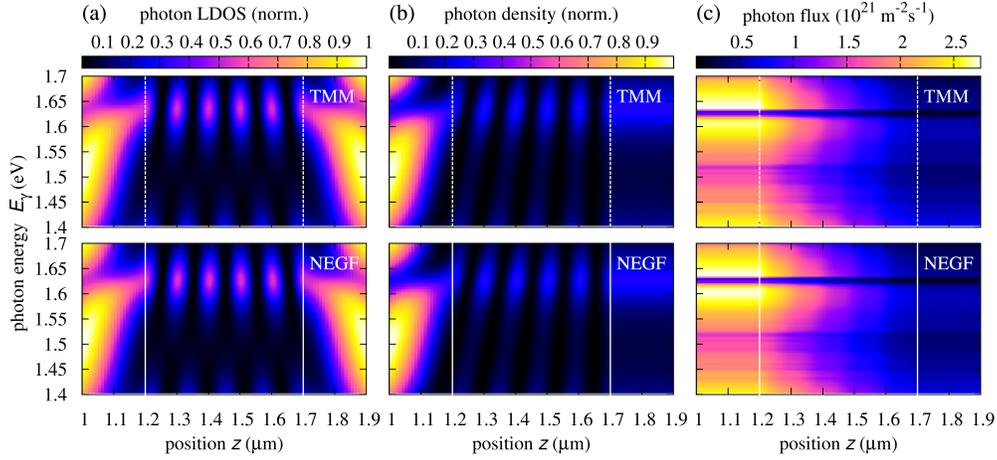}
\caption{Comparison between the results provided by the transfer
matrix method (TMM) and the non-equilibrium Green's function formalism (NEGF) for (a) the
local density of photon states at $\mathbf{q}_{\parallel}=0$, (b) the local photon density for
normal incindence from the left, and (c) the local photon flux (modal Poynting vector) under
illumination with the solar AM1.5g spectrum of a 500 nm GaAs slab.\label{fig:tmm_negf}}
\end{center}
\end{figure} 
For the LDOS, the NEGF expression \eqref{eq:ldos} is compared to the sum of the absolute value
squared of the electric field for unit left and right incidence in TMM, 
$\mathcal{N}^{TMM}(z,E)\propto |\mathcal{E}(z,E)|^{2}E^{2}$ [Fig. \ref{fig:tmm_negf}(a)], with
normalization to the maximum value. In the regions left and right of the semiconductor slab, the
numerical value of the LDOS at $q_{\parallel}=0$ approaches that of the analytical expression 
$\mathcal{N}_{0}^{(1)}(q_{\parallel}=0,z,E)=n_{0}/\left(\pi\hbar c_{0}A\right)$ for
homogeneous bulk, which is reproduced exactly for a constant real index of refraction.
For the photon density, the numerical value of expression \eqref{eq:photdens} with the correlation
function resulting from the solution of \eqref{eq:keldysh} under the assumption of an asymmetric mode occupation 
$N_{\mathbf{q}}=\tilde{N}(E)\delta(q_{\parallel})\theta(q_{z})$ is compared to the absolute value of
the electric field for left incidence in TMM [Fig. \ref{fig:tmm_negf}(b)], again normalized to the
maximum value. The decay of the photon mode occupation is monitored via the Poynting vector. Using
the expression for the non-interacting GFs \eqref{eq:corrfn_free_in} and  \eqref{eq:corrfn_free_out}
at 	$\mathbf{q}_{\parallel}=0$ in \eqref{eq:poyntvec} fixes the mode occupation $\tilde{N}(E)$ for a
given 	incident photon flux: for normal incidence from the left, the only non-vanishing components
of the 	correlation functions for $E>0$ are 
\begin{align}
\mathcal{D}_{0\mu\mu}^{<}(\mathbf{0},z,z',E)=&-\frac{in_{0}^{2}}{2A}\exp[iq_{0}(z-z')]q_{0}^{-1}\tilde{N}(E),\\
\mathcal{D}^{>}_{0\mu\mu}(\mathbf{0},z,z',E)=&\frac{in_{0}^{2}}{2A}\Big\{\exp[iq_{0}(z-z')]\tilde{N}(E)
+2\cos[q_{0}(z-z')]\Big\}q_{0}^{-1},\qquad
\mu=x,y.
\end{align}
The second term in $\mathcal{D}^{>}_{0}$ originates from spontaneous emission due to vacuum
fluctuations. However, this term gives no contribution to the Poynting vector, since it is entirely
imaginary. The remaining expression then amounts to
\begin{align}
s_{z}(\mathbf{q}_{\parallel},z,E)=&-\frac{E}{2\pi\hbar}\lim_{z'\rightarrow
z}\partial_{z'}\Big\{\frac{n_{0}^{2}}{A}\sin[q_{0}(z-z')]\tilde{N}(E)\Big\} 
=\frac{n_{0}^{2}E}{2\pi\hbar A}\tilde{N}(E).
\end{align}
In terms of the incident modal intensity $s_{0z}$ (with units of photon flux), the mode
occupation is thus given by
\begin{align} 
\tilde{N}(E)=\frac{s_{0z}(E)2\pi\hbar A}{n_{0}^{2}|E|}
\end{align} 
 Fig. \ref{fig:tmm_negf}(c) shows the numerical values for a photon flux corresponding to the
 fraction of the AM1.5g solar spectrum in the range 1.4-1.7 eV. For all the cases compared, and inspite of the
NEGF being a numerical method while the TMM is semi-analytical, the accuracy is remarkable.

\begin{figure}[tb]
\begin{center} 
\includegraphics[width=\textwidth]{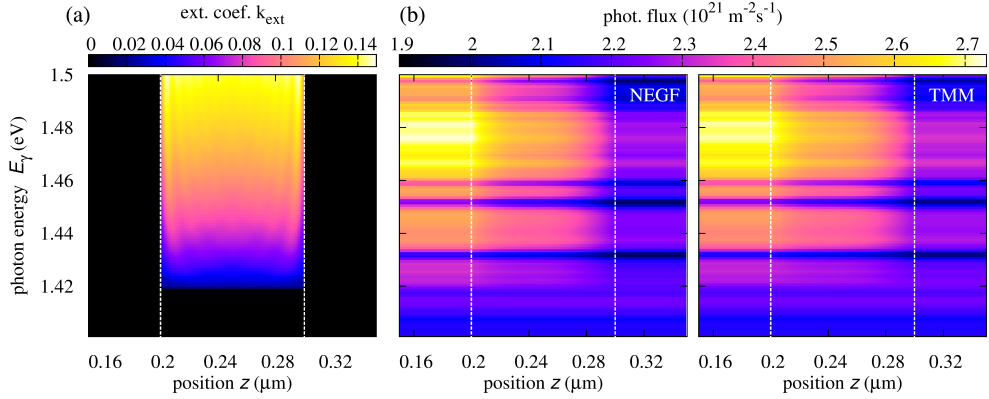}
\caption{(a) Local extinction coefficient $\kappa(z)$ from the spatial
average of the non-local photon self-energy, (b) $z$-component of the modal Poynting vector from
NEGF with non-local coupling via the full self-energy and from the TMM with a spatially averaged
extinction coefficient.
\label{fig:abs_negf_nonloc}}
\end{center}
\end{figure}

Finally, the light propagation for non-local coupling is compared to that for local coupling. For
the description of the non-local response of the semiconductor slab, the expression
\eqref{eq:photse} for the photon self-energy is evaluated using the analytical form of the slab GF
for a two-band effective mass model of a homogeneous semiconductor. Only the imaginary part of the retarded and
greater self-energy components is used, and, for a more realistic description, the real refractive
index profile is substituted for the diagonal entries of the real part of the self-energy; also, the
self-energy is assumed to be diagonal in the polarization index. Fig. \ref{fig:abs_negf_nonloc}(a)
shows the corresponding local extinction coefficient according to \eqref{eq:extcoef} for a 100 nm
slab in the energy range 1.4-1.5 eV, while the photon flux computed using the fully non-local
photon 	self-energy is given in Fig. \ref{fig:abs_negf_nonloc}(b), together with the TMM result for 
the photon flux obtained with this local extinction coefficient using a spatial average 
$\bar{\kappa}=(z_{max}-z_{0})^{-1}\int_{z_{0}}^{z_{max}} dz\kappa(z)$. As can be inferred from these results, the approximation of
local 	coupling tends to slightly underestimate the overall absorption, however, the deviation is in
the range of the inaccuracy due to the numerical solution of equations \eqref{eq:dyson} and \eqref{eq:keldysh}.

\section{Conclusions} 
In this paper, a theoretical description and numerical solution of the equations governing the
propagation of light in layered media was provided on the level of quantum statistical mechanics for
an arbitrary non-equilibrium steady state and non-local coupling between photonic modes and
electronic states. The formalism was extensively verified for normal incidence on homogeneous slab
systems with spatially averaged local electron-photon interaction, finding excellent agreement with
the results provided by the conventional semi-analytical transfer matrix method. Application of the formalism to
the case of fully non-local coupling revealed only a minimal deviation from the predictions of
the local approximation.

\section*{Acknowledgements}

The author would like to acknowledge the support and kind hospitality of the National Renewable
Energy Laboratory in Golden, Colorado, USA, during his visit in the framework of the
Helmholtz-NREL Joint Research Initiative HNSEI. Financial support was provided in part by the
European Union FP-7 Programme via grant No.~246200.

\appendix

\section*{Appendix}

\section{Noninteracting GF\label{sec:appa}}

The special case of  the \emph{free} photon GF for homogeneous bulk in \emph{equilibrium} is
described by
\begin{align}
\mathcal{D}_{0\mu\nu}(1,2)=\sum_{\mathbf{q}}e^{i\mathbf{q}(\mathbf{r}_{1}-\mathbf{r}_{2})}\mathcal{D}_{0\mu\nu}(\mathbf{q};t_{1},t_{2}),
\end{align}
with
\begin{align}
\mathcal{D}_{0\mu\nu}(\mathbf{q};t,t')=&\frac{\hbar
c_{0}^{2}}{2V}\sum_{\lambda}(\omega_{\lambda\mathbf{q}})^{-1}\Big[{\mathbf
\epsilon}_{\lambda{\mathbf q}}^{\mu}{\mathbf\epsilon}_{\lambda{\mathbf
q}}^{\nu}D_{0\lambda}^{'}(\mathbf{q};t,t')+{\mathbf \epsilon}_{\lambda(-{\mathbf q})}^{\mu}{\mathbf\epsilon}_{\lambda(-{\mathbf
q})}^{\nu}D_{0\lambda}^{''}(-\mathbf{q};t,t')\Big]\nonumber\\
=&\frac{\hbar c_{0}^2}{2V}\sum_{\lambda}\frac{{\mathbf 
\epsilon}_{\lambda{\mathbf q}}^{\mu}{\mathbf\epsilon}_{\lambda{\mathbf
q}}^{\nu}}{\omega_{\lambda\mathbf{q}}}\Big[D_{0\lambda}^{'}(\mathbf{q};t,t')+D_{0\lambda}^{''}(-\mathbf{q};t,t')\Big]\\
\equiv&\frac{\hbar c_{0}^2}{2V}\sum_{\lambda}\frac{{\mathbf 
\epsilon}_{\lambda{\mathbf q}}^{\mu}{\mathbf\epsilon}_{\lambda{\mathbf
q}}^{\nu}}{\omega_{\lambda\mathbf{q}}}D_{0\lambda}(\mathbf{q};t,t'),
\end{align}
where the bare scalar photon (boson)
propagator is defined as  
\begin{align}
D_{0\lambda}(\mathbf{q};t,t')\equiv& -\frac{i}{\hbar}\left\langle
\hat{T}_{C}\left\{[\hat{b}^{\dagger}_{\lambda,-\mathbf{q}}(t)+
\hat{b}_{\lambda,\mathbf{q}}(t)]
[\hat{b}^{\dagger}_{\lambda,\mathbf{q}}(t')+\hat{b}_{\lambda,-\mathbf{q}}(t')]\right\}\right\rangle_{0}\\
=&-\frac{i}{\hbar}\left[\left\langle
\hat{T}_{C}\left\{\hat{b}_{\lambda,\mathbf{q}}(t)\hat{b}^{\dagger}_{\lambda,\mathbf{q}}(t')\right\}\right\rangle_{0}+\left\langle
\hat{T}_{C}\left\{\hat{b}^{\dagger}_{\lambda,-\mathbf{q}}(t)\hat{b}_{\lambda,-\mathbf{q}}(t')\right\}\right\rangle_{0}\right].
\end{align}
The corresponding real-time steady-state expressions are 
\begin{align}
  D_{0\lambda}^{\lessgtr}({\mathbf q},E) &=-2\pi i\left[N^{0}_{\lambda,{\mathbf 
  q}}\delta(E\mp \hbar\omega_{\lambda{\mathbf q}})+(N^{0}_{\lambda,-{\mathbf
        q}}+1)\delta(E\pm\hbar\omega_{\lambda{\mathbf
        q}})\right],\label{eq:freephotbulkprop}\\ D_{0\lambda}^{R,A}({\mathbf
        q},E)&=\frac{1}{E-\hbar\omega_{\lambda{\mathbf q}}\pm i\eta}-\frac{1}{E+\hbar\omega_{\lambda{\mathbf q}}\pm
        i\eta}\\
        &=\frac{2\hbar\omega_{\lambda{\mathbf q} }}{(E+i\eta)^2-(\hbar\omega_{\lambda{\mathbf q}
        })^2},
\label{eq:equilphonen}
\end{align}
where $\eta\rightarrow 0^{+}$ and
\begin{align}
N_{\lambda,{\mathbf q}}^{0}\equiv&
\langle\hat{b}^{\dagger}_{\lambda,\mathbf{q}}\hat{b}_{\lambda,\mathbf{q}}\rangle_{0}
\end{align}
is the occupation of photon mode $(\lambda,\mathbf{q})$.

\section{Dyadic propagator in cylindrical coordinates\label{sec:appb}}
For a situation with spatial isotropy in the transverse dimensions, cylindrical
coordinates can be used, where
$\mathbf{q}_{\parallel}=q_{\parallel}\left(\cos\varphi,
\sin\varphi\right)$,
and the azimuthal dependence of the GF can be separated from the
dependence on the absolute value of the transverse momentum, i.e.,
$\mathcal{D}_{0\mu\nu}(\mathbf{q}_{\parallel},z,z',E)=
\mathcal{D}_{0\mu\nu}(q_{\parallel},\varphi,z,z',E)$.
In this case, the full retarded GF can be written as  
\begin{align}
\mathcal{D}_{0\mu\nu}^{R}(q_{\parallel},\varphi,z,z',E)=\mathcal{H}(q_{\parallel},z,z',E)
\mathcal{F}_{\mu\nu}(q_{\parallel},\varphi,z,z',E)
\end{align}
with the scalar, angle-independent Huygens propagator
\begin{align}
\mathcal{H}(q_{\parallel},z,z',E)=-\frac{in_{0}^2}{2A}\frac{\exp\left[iq_{z0}(q_{\parallel},E)|z-z'|\right]}
{q_{z0}(q_{\parallel},E)}
\end{align}
and 
\begin{align}
&\mathcal{F}_{\mu\nu}(q_{\parallel},\varphi,z,z',E)=q_{0}^{-2}\nonumber\\
&\times\begin{pmatrix}
q_{0}^{2}-q_{\parallel}^{2}f_{xx}(\varphi)&-q_{\parallel}^{2}f_{xy}(\varphi)&-q_{\parallel}q_{z0}f_{xz}(\varphi)\mathrm{sgn}(z-z')\\
-q_{\parallel}^{2}f_{xy}(\varphi)&q_{0}^{2}-q_{\parallel}^{2}f_{yy}(\varphi)&-q_{\parallel}q_{z0}f_{yz}(\varphi)\mathrm{sgn}(z-z')\\
-q_{\parallel}q_{z0}f_{xz}(\varphi)\mathrm{sgn}(z-z')&-q_{\parallel}q_{z0}f_{yz}(\varphi)\mathrm{sgn}(z-z')&q_{\parallel}^{2}
\end{pmatrix},
\end{align}
where the angular factors $f_{\mu\nu}(\varphi)$ are given by
\begin{align}
f_{xx}(\varphi)=\cos^2\varphi,~f_{yy}(\varphi)=\sin^2\varphi,~f_{zz}(\varphi)=1,\\
f_{xy}(\varphi)=\sin\varphi\cos\varphi,~f_{xz}(\varphi)=\cos\varphi,~f_{yz}(\varphi)=\sin\varphi.
\end{align}
In the continuum limit, the summation over transverse photon momentum in the expressions for LDOS
[Eq. \eqref{eq:ldos}] and local photon density [Eq. \eqref{eq:photdens}] is replaced by integrations
over absolute value of momentum and over angle, 
\begin{align}
\sum_{\mathbf{q}_{\parallel}}\rightarrow\frac{A}{(2\pi)^{2}}\int_{0}^{q_{0}}
dq_{\parallel}q_{\parallel}\int_{0}^{2\pi} d\varphi
\end{align}
Upon the angular integration, only the diagonal terms of $\mathbf{f}$ survive
($\bar{f}_{\mu\nu}=\int d\varphi f_{\mu\nu}$):
\begin{align}
\bar{f}_{xx}=\bar{f}_{yy}=\pi,\quad \bar{f}_{zz}=2\pi. 
\end{align}
The angular average of the diagonal anisotropic components of the retarded GF for free field modes
is thus given by
\begin{align}
\bar{\mathcal{D}}_{0xx}^{R(2)}(q_{\parallel},z,z,E)=&\bar{\mathcal{D}}_{0yy}^{R(2)}(q_{\parallel},z,z,E)\\
=& -\pi q_{\parallel}^{2}q_{0}^{-2}
\mathcal{H}(q_{\parallel},z,z,E)=\frac{in_{0}^2}{2A}\frac{\pi
q_{\parallel}^{2}}{q_{z0}(q_{\parallel},E)q_{0}^{2}},\\
\bar{\mathcal{D}}_{0zz}^{R(2)}(q_{\parallel},z,z,E)=&-2\pi q_{z0}(q_{\parallel},E)q_{0}^{-2}
\mathcal{H}(q_{\parallel},z,z,E)=\frac{in_{0}^2}{2A}\frac{2\pi
q_{z0}(q_{\parallel},E)}{q_{0}^{2}}.
\end{align}
Integration over transverse momentum und normalization with $\mathcal{C}=\frac{2E}{(\hbar
c_{0})^{2}}$ provides the anisotropic DOS components via 
\begin{align}
\mathcal{N}_{\mu\mu}^{(2)}(z,E)=&-\frac{\mathcal{C}}{\pi}\frac{A}{(2\pi)^{2}}\int_{0}^{q_{0}}
dq_{\parallel}q_{\parallel}
\mathrm{Im}\bar{\mathcal{D}}_{\mu\mu
z}^{R(2)}(q_{\parallel},z,z,E)\\
=&-\frac{1}{3}\frac{n_{0}^{3}E^{2}}{2\pi^{2}(\hbar
c_{0})^{3}},\quad \mu=x,y,z.
\end{align}
The isotropic part of the angle-averaged diagonal retarded GF is
\begin{align}
\bar{\mathcal{D}}_{0\mu\mu}^{R(1)}(q_{\parallel},z,z,E)=&2\pi
\mathcal{H}(q_{\parallel},z,z,E)=-\frac{in_{0}^2}{2A}\frac{2\pi}{q_{z0}(q_{\parallel},E)},\quad
\mu=x,y,z,
\end{align}
which provides the isotropic part of the LDOS components
\begin{align}
\mathcal{N}_{\mu\mu}^{(1)}(z,E)
=&\frac{n_{0}^{3}E^{2}}{2\pi^{2}(\hbar
c_{0})^{3}},\quad \mu=x,y,z.
\end{align}

\end{document}